%% file: transfer_learning_protein_quality_preprint.tex
\documentclass[aps, pra, a4paper, twocolumn, floatfix]{revtex4-1}

\usepackage[utf8]{inputenc}
\usepackage[T1]{fontenc}
\usepackage{lmodern}

\usepackage{graphicx}
\usepackage{hyperref}       
\usepackage{url}            
\usepackage{booktabs}       
\usepackage{amsfonts}       
\usepackage{nicefrac}       

\usepackage{amsmath}
\usepackage{amssymb}
\usepackage{amsthm}
\usepackage{subcaption}
\usepackage{float}
\usepackage{siunitx}

\usepackage{placeins}
\usepackage{natbib}

\usepackage{booktabs}

\begin{document}
\title{Deep transfer learning in the assessment of the quality of protein models}
\author{David \surname{Menéndez Hurtado}}
\email{david.menendez.hurtado@scilifelab.se}
\affiliation{Department of Biochemistry and Biophysics and Science for Life Laboratory, Stockholm University}

\author{Karolis Uziela}
\affiliation{Department of Biochemistry and Biophysics and Science for Life Laboratory, Stockholm University}

\author{Arne Elofsson}
\email{arne@bioinfo.se}
\affiliation{Department of Biochemistry and Biophysics and Science for Life Laboratory, Stockholm University}

		\begin{abstract}
			\textbf{Motivation:} Proteins fold into complex structures that are crucial for their biological functions.
			Experimental determination of protein structures is costly and therefore limited to a small fraction of all known proteins.
			Hence, different computational structure prediction methods are necessary for the modelling of the vast majority of all proteins.
			In most structure prediction pipelines, the last step is to select the best available model and to estimate its accuracy.
			This model quality estimation problem has been growing in importance during the last decade, and progress is believed to be important for large scale modelling of proteins.
			The current generation of model quality estimation programs performs well at separating incorrect and good models, but fails to consistently identify the best possible model.
			State-of-the-art model quality assessment methods use a combination of features that describe a model and the agreement of the model with features predicted from the protein sequence.\\
			\textbf{Results:} We first introduce a deep neural network architecture to predict model quality using significantly fewer input features than state-of-the-art methods.
			Thereafter, we propose a methodology to train the deep network that leverages the comparative structure of the problem.
			We also show the possibility of applying transfer learning on databases of known protein structures.
			We demonstrate its viability by reaching state-of-the-art performance using only a  reduced set of input features and a coarse description of the models.\\
			\textbf{Availability:} The code will be freely available for download at \url{github.com/ElofssonLab/ProQ4}.\\
		\end{abstract}
\maketitle
\section{Introduction}

Proteins perform the vast majority of all biological functions.  They
are constructed as long polymers built of twenty amino acids that fold
into a three-dimensional shape after synthesis.  During the last half
of the century, the structures of more than 100 000 proteins have been
experimentally obtained using X-ray crystallography, NMR, or electron
microscopy.  Although the experimental techniques have improved
significantly, the average cost for a new protein structure is still
close to \$100k, limiting the number of experimentally determined
protein structures~\citep{Terwilliger19416074}.

A single organism contains thousands of genes, each coding for at least
one protein. Due to the exponential decrease of sequencing costs the
sequence of more than 100 million proteins has been deposited in
public databases like UniRef~\citep{uniprot}.
Further, almost one order of magnitude more sequences are
available from meta-genomic projects. This means that there are at
least three orders of magnitude separating the number of known protein
sequences and structures. This gap is increasing, and only
computational methods will be able to close it.

Luckily, methods to generate accurate three-dimensional models of
proteins exist. If the structure of a homologous protein has been
solved, it can be used as a template for modelling. This
method can be applied to about 50\% of all protein families \citep{pconsfold2, bakerfold}.
For the rest, one has to rely on other methods.
Here, rapid progress in contact prediction
has recently enabled the modelling of the structure of many proteins. The
accuracy of these models varies and depends on several factors and no
single method always produces the best model.  Therefore, in many
modelling approaches a combination of methods and/or parameter settings
are used to produce multiple models.
This will later will be analysed to identify the best one.

\subsection{Deep learning}
Deep learning is a family of machine learning algorithms that has brought a revolution in fields such as computer vision, speech recognition, and artificial intelligence~\citep{deeplearning}.
The improvements can be attributed to two discoveries: (i) new algorithms enable to define and efficiently train much more complex models, that can take advantage of large training sets; and (ii), deep learning methods can make use of the structure of the data to take a data-driven approach to feature engineering.
Instead of hand crafting high-level features, deep learning can use lower lever representations of the data that preserve its structure, such as the ordering of amino acids in a protein chain, or the proximity of pixels in an image.
A deep learning model is composed of multiple layers that successively transform the inputs, learning to extract the relevant information during training.
In effect, replacing the manual feature engineering with an automatic feature extraction.
In a deep model, each layer learns to combine some of its input creating a hierarchy of features automatically extracted from the data.
For example, when trying to identify an object in an image, the first layers might learn to detect edges, the next layers would combine them into textures, object parts, and eventually the whole training label.

\subsection{Transfer learning}
Deep learning can be used to train high capacity models, but it requires large amounts of data, that is not always available or cheap.
But since most of the network acts as a feature extractor, it is possible to train a network on a larger and different but related problem, and reuse the learned feature extraction.

\citet{transfer_features} studied how well these features are generalisable across categories in image recognition and has since become a standard procedure in deep learning.

\subsection{Predicting protein structural features}
The structure of a protein consists of secondary structure elements;
$\alpha$-helices and $\beta$-sheets that are packed in such a way that
hydrophobic amino acids are mostly hidden from the surface.  We
also know that homologous sequences fold into similar structures, so we
can search sequence databases for proteins that are likely to be
similar to the one we are interested in, and compile them into an
aligned list called multiple sequence alignment.

Both the secondary structure and the surface accessibility of a
residue in a protein can be predicted with a rather high accuracy by
applying machine learning on the statistics extracted from the columns
of this multiple sequence alignment.  The accuracy of these methods
has reached close to 80\% in two \citep{netsurf} or three \citep{psipred} state classifications.

\subsection{CASP: Critical Assessment of protein Structure Prediction}

CASP is a biennial experiment aimed at assessing the state of the
structure prediction field.
The organizers release the sequence of proteins of hitherto unknown structure, and allow a few days for groups around the world to submit predictions.
In a second stage, these models are published and evaluated by model quality assessment methods.
The latest editions count around 100 individual sequences (also called targets), and received around 200 models per target, coming from circa 30 independent methods.

\subsection{Model quality assessment background}
Estimating the free energies of protein models has a long history
within the protein structure prediction field.~\citep{park_energy, lazaridis_energy}
Here, the ultimate goal is to understand the fundamental physics governing protein folding to such an extent that it is possible to simulate the folding of a protein.
The purpose of the model quality assessment program is to accurately estimate the distance of the model from the native structure.

There are two general strategies to evaluate the quality of a single
protein model: comparison with auxiliary predictions, and evaluation
of physico-chemical properties of the model. A limitation of the second
approach is be that even if a method could
perfectly describe the free energy of a protein model - this measure
might not be at all correlated with the difference of a model from the
native structure. Take the trivial example of a model where all atoms
except one, are perfectly placed. This last atom is then positioned on top
of another atom. Such a model would have a near infinite free
energy, but by any distance measure, be almost perfect. 

\begin{figure}[t]
 \centering
 \includegraphics[width=0.9\linewidth]{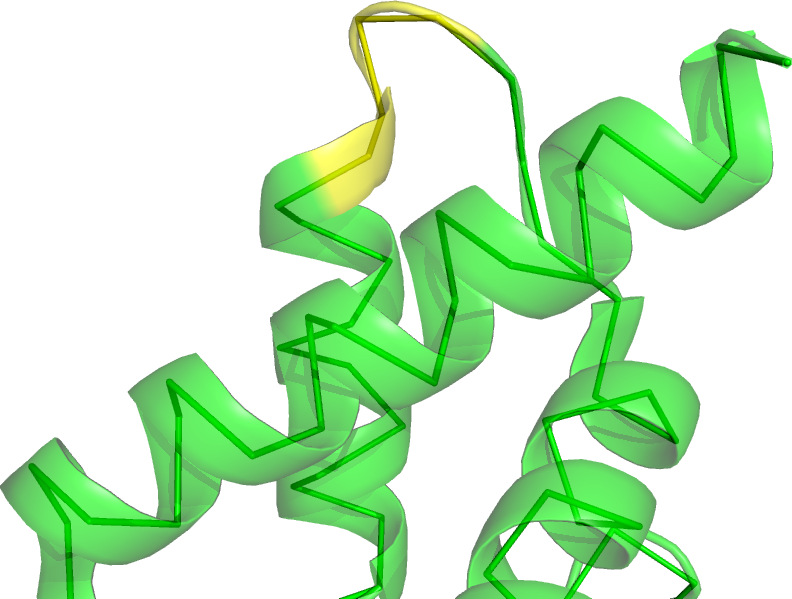}
 \caption{Detail of the 3D structure of the protein 3TDU. Highlighted in yellow are the residues that smoothly transition between helix and coil. Predictions are commonly wrong about the exact position of the boundary.}\label{subfig:ss_3d}
 \label{fig:ss}
\end{figure}

On the other hand models solely relying on predictions might provide
important low-resolution methods and reduce the complexity of 3D
prediction to simpler ones, such as secondary structure.
A good model is expected to agree with the predictions. In addition, all disagreements might not carry the same
information. Often the exact boundaries of secondary structure elements are
notoriously difficult to predict, and depend on the exact
thresholds used to define it, as seen in Figure~\ref{fig:ss}. In
contrast, it is rare that errors in predictions between different
secondary structure classes occur.

\subsection{Related work}
For more than a decade several groups have developed protein model
quality assessment methods using various input
features~\citep{ProQ,ProQres,ProQ2,Mirzaei28113636,svmqa,modfold6}.
The improvements achieved during the last few years can be attributed
to small, but significant, improvements to the methods by (i) including
additional descriptions of a protein model~\citep{ProQ3}, (ii) applying
deep learning techniques~\citep{ProQ3D} and combining many
features~\citep{modfold6}.

Here follows a brief description of the inputs and algorithms used by other comparable methods.
These mostly focus on physico-chemical descriptors, such as statistical potentials, and train simple machine learning algorithms.
\begin{itemize}
	\item \emph{QMEAN} \citep{qmean}: Linear combination of the agreement of secondary structure and surface area, and three statistical potentials describing the torsion angles, amino acid contacts, and exposure to the water.
	\item \emph{DeepQA} \citep{deepqa}: Deep Belief Network combining several scoring potentials and energies, other quality assessment methods, and seven different agreement metrics between observed and predicted secondary structure and surface area.
	\item \emph{ProQ3} \citep{ProQ3}: A Linear SVM trained on atom and amino acid contacts,
	observed secondary structure, multiple sequence alignment
	statistics, agreements with secondary structure and surface area, as
	well as terms of the Talaris energy function~\citep{talaris}. 
	\item \emph{ProQ3D} \citep{ProQ3D}: Same input as ProQ3, but replacing the linear SVM with a multi layer perceptron.
	\item \emph{VoroMQA} \citep{voromqa}: Statistical potential based on the frequencies of observed atom contacts.
	\item \emph{SVMQA} \citep{svmqa}: SVM combining different statistical potentials and agreement with secondary structure and surface area.
\end{itemize}

In contrast to all other methods, our method uses only coarse structural features and a multiple sequence alignment, but no statistical potentials nor any other chemical description.

\section{Methods}

\subsection{Datasets}
In order to directly compare results with the most recent method,
ProQ3D~\citep{ProQ3D} retrained on LDDT~\citep{target_function}, we have used the same datasets for training
and testing: all the submitted models to the CASP editions 9 and 10
for training, and CASP 11 for evaluation, excluding all the targets shorter
than 50 residues.
In total, we have \num{57263} models from \num{212} different targets in the training set, and \num{14580} from 79 targets in the test set, excluding targets cancelled by the organisers.

We also present the results on the same subset of the Cameo dataset used by ProQ3D, \num{19899} models from 676 different targets.

For the pre-trained networks, we used \num{5687} structures from the
PISCES dataset~\citep{pdbcull}, while an additional 300 were left out for validation.
The multiple sequence alignments were produced using Jackhmmer~\citep{jh} to search Uniref50~\citep{uniref}, with an E-value threshold of
$10^{-3}$ for 3 iterations.

\subsection{Description of inputs}
Prediction of structural features of a protein is improved by using multiple sequence alignments~\citep{redefining_ss}.
From the multiple sequence alignment we extract two statistics: the self-information (Equation~\ref{eq:selfinfo}) and the partial entropy
(Equation~\ref{eq:partentro}) of the position:

\begin{subequations}
	\begin{equation}
		I_i=-\log\left(\frac{p_i}{\bar{p_i}}\right), \label{eq:selfinfo}
	\end{equation}
	\begin{equation}
		S_i=-p_i \log\left(\frac{p_i}{\bar{p_i}}\right), \label{eq:partentro}
	\end{equation}
\end{subequations}
where $p_i$ is the frequency of the amino acid $i$ at the position and
$\bar{p_i}$ is the average frequency on the data set.  We also include the protein sequence itself using sparse
encoding.

The protein models are defined in a coarse representation by the sine and cosine of the dihedral angles~\citep{Xue18214956}
$\varphi$ and $\psi$, the secondary structure, relative
surface area, and energies of hydrogen bonds in the backbone, as defined by DSSP~\citep{dssp}.
In DSSP, 8 different secondary structure states are assigned to a
protein. Due to lack of data, we merged the following two pairs of states:  G
($3_{10}$ helix) and I ($\pi-$ helix), and T (hydrogen bonded turn)
and S (bend).

\subsection{Output: the target function}
There are several different metrics that can be used to evaluate the similarity between a protein model and the native structure.
In this work we choose to use a local scoring function: the Local Distance
Difference Test (LDDT)~\citep{lddt}, a number between $0$ and $1$ that
represents the fraction of conserved contacts between all pairs of
atoms in the native structure for several distance thresholds. An LDDT
of $1$ means a perfect agreement, and a good model shall have scores
higher than $0.5$. Other quality estimation methods could also be used
providing similar performance.
To estimate the quality of a model we define the global score to be
the average of the local scores, giving a score of $0$
to any residue missing from the model, and ignoring those absent from the native
structure.

\subsection{Figures of merit}\label{sec:metrics}
The performance of the models will be evaluated on several metrics.
\begin{itemize}
	\item \emph{Local correlation:} Pearson correlation between the predicted and true values assigned to every residue. This measures the reliability of the predicted scores assigned to each residue.
	\item \emph{Local RMSE:} Root Mean Squared Error between the aforementioned predicted local scores and the true values.
	\item \emph{Per model correlation:} average Pearson correlation between predicted and true values of every residue for every model. This measures how well we can differentiate the correct and incorrect parts of the model.
	\item \emph{Global correlation:} Pearson correlation between predicted and true values for the overall model. This is defined as the average of the local scores, ignoring any residue not present in the native structure, and setting the score of missing residues to 0. This measures how well we can rank targets in a global scale.
	\item \emph{Global RMSE:} Root Mean Squared Error between the global predicted and true scores.
	\item \emph{Per target correlation:} average Pearson correlation for the global scores per target. A measurement of the ability of the program to rank models from the same target.
	\item \emph{First rank loss:} average difference between the best and the top ranked model. It measures how well we can select the best model for each target.
\end{itemize}

\section{Implementation}
\subsection{Network architecture}
In order to exploit the spatial distribution of the local features,
and to allow the network to compare observations and predictions
locally, we have implemented a 1D fully convolutional network trained on the local scores. 
We used the
ELU activation function~\citep{elu} as a non-linearity,
and also applied a small $L^2$ penalty of $10^{-14}$ to every convolutional
layer, except for the output layers, 
The training was guided by the
Adam optimiser~\citep{adam}.

\subsection{The ResNet module}
\begin{figure}[h]
	\centering
	\includegraphics[width=0.4\linewidth]{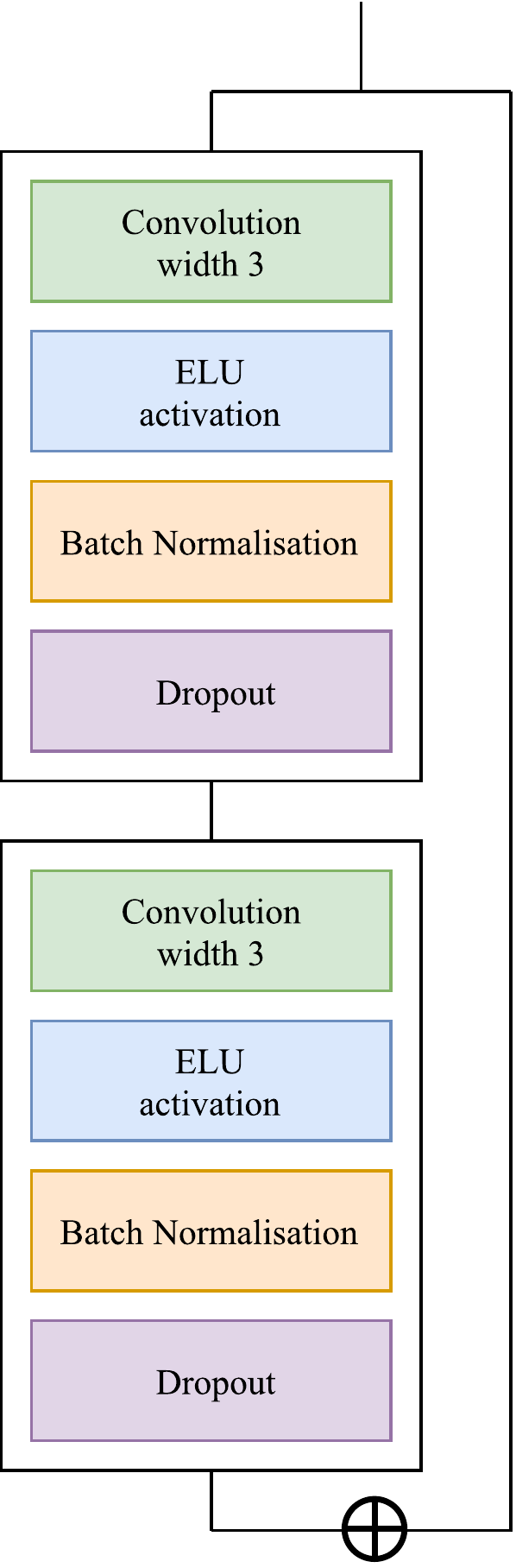}
	\caption{The 1D ResNet module, the main building block of our convolutional nets}
	\label{fig:resnet}
\end{figure}

The convolutional block from ResNet~\citep{resnet} inspired our
architecture due to its capacity to converge
efficiently while still keeping a deep architecture that provides a large
effective field of view. 
As shown on Figure~\ref{fig:resnet}, it is composed of two blocks of successive convolution of width 3, followed by an activation
function, batch normalisation~\citep{bn}, and dropout, whose output is summed to the inputs.
When the number of input and output channels is different, we modify the skip connection to include only one convolution, activation, batch normalisation, and dropout.

\subsection{Simple Convolutional Neural Network}
\begin{figure}
	\centering
	\includegraphics[width=\linewidth]{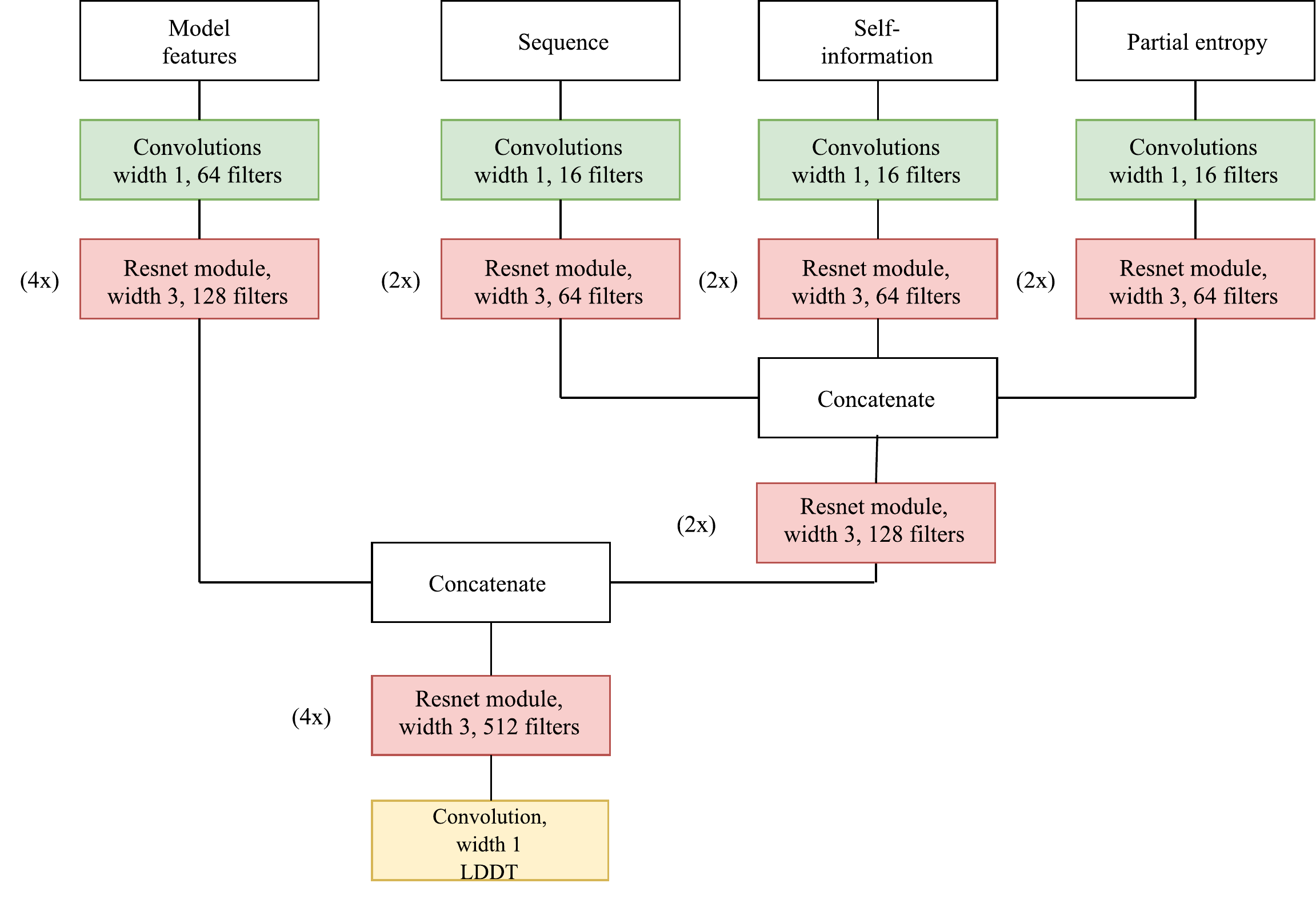}
	\caption{The convolutional architecture}
	\label{fig:cnn}
\end{figure}

The simplest implementation of a deep convolutional network consists of several branches that combines all the inputs can be seen in Figure~\ref{fig:cnn}.
In one branch, the three input vectors derived from the sequence (the sequence itself, self-information, and partial entropy) are followed by a convolution of size 1, to project them into a 16 dimensional vector space per residue.
They are then followed by two ResNet modules, merged into a single branch by concatenation, and followed by two more ResNet modules.
The structural inputs are likewise projected into a 64 dimensional space and passed through four ResNet modules.
Finally, both branches are concatenated, and sent through four more ResNet modules.
A single convolutional layer of width 7 and no $L^2$ penalty is
applied used in the final layer.

\subsection{Sequence pre-trained network}
Our dataset contains roughly 200 times more models than unique sequences, which is an obvious source of bias.
For example, some of these targets are hard, and not a single model is of good quality.
The network could therefore easily learn that this particular sequence is always bad, which we want to avoid.
After all, a new method, or more data could result in a good model in the future.

\begin{figure*}
	\centering
	\begin{subfigure}{0.45\linewidth}
		\includegraphics[width=0.9\linewidth]{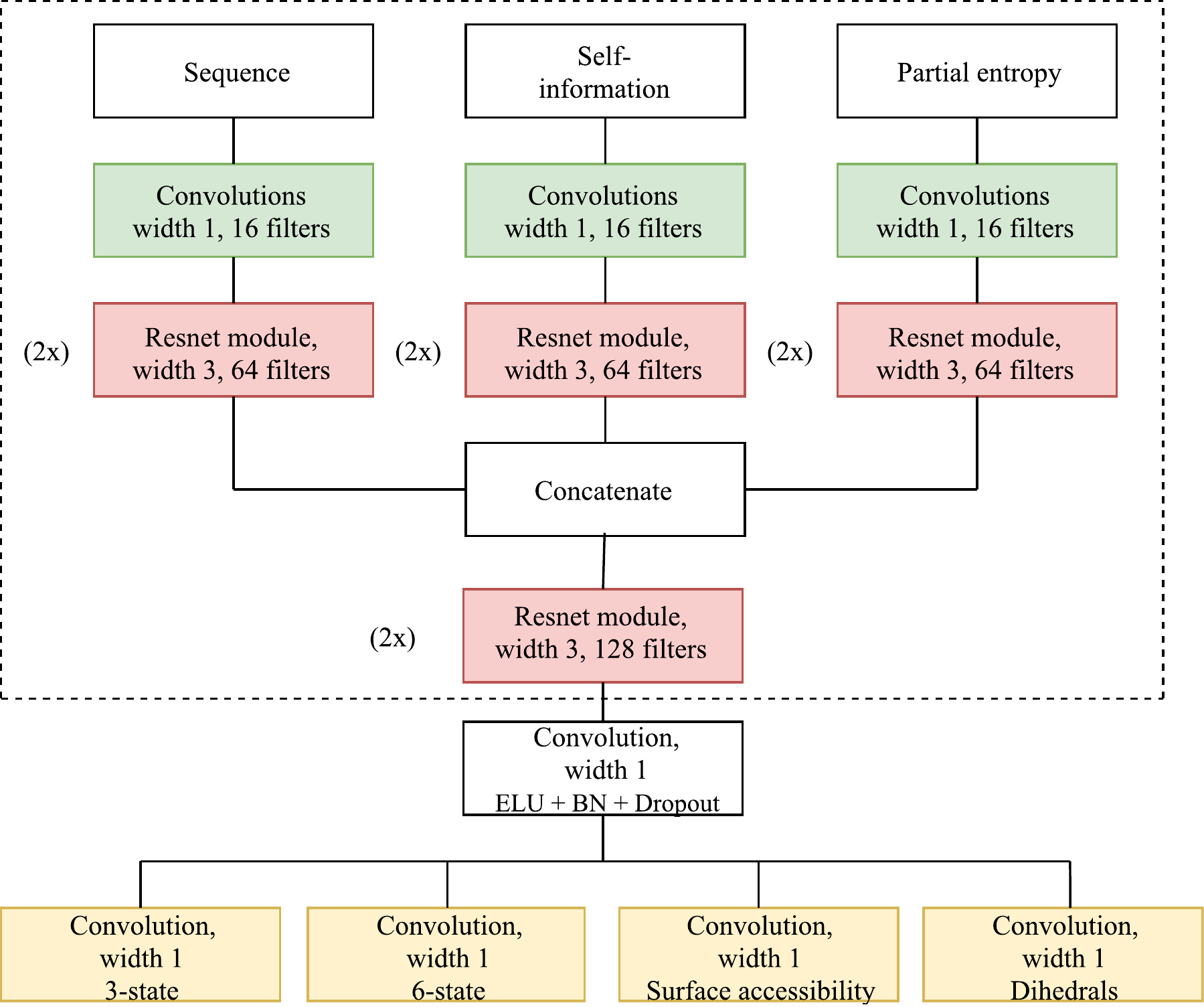}
		\caption{Sequence pre-training, learning the alignment features.}
		\label{subfig:seq}
	\end{subfigure} %
	\begin{subfigure}{0.45\linewidth}
		\includegraphics[width=0.9\linewidth]{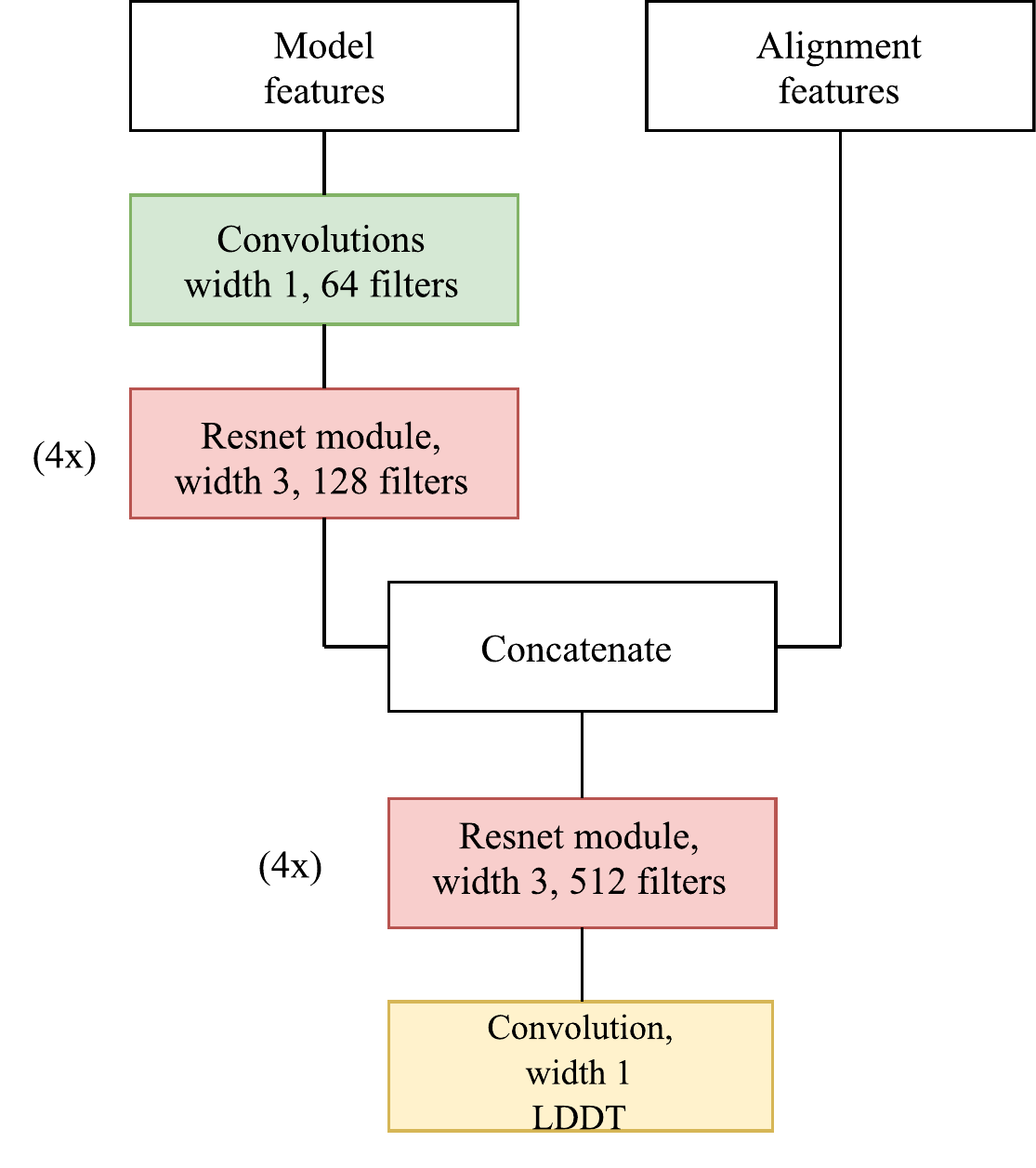}
		\caption{Sequence pre-trained network}
		\label{subfig:pretrained}
	\end{subfigure}
	\caption{The two stages of the Pre-trained network. The sequence pre-training is used to extract the alignment features that are used subsequently throughout the paper.}
	\label{fig:pretrained}
\end{figure*}

To tackle this issue, we pre-train the branch corresponding to the sequence inputs on \num{5687} of known protein structures from PISCES~\citep{pdbcull}, showing our network a much more diverse set of proteins.
As shown in Figure~\ref{subfig:seq}, we inserted one hidden convolutional layer connected to four more predicting the 3 and 6-state secondary structure, surface accessibility, and sine and cosine of the dihedral angles.
We denominate \emph{alignment features} to the output of the network in the hidden layer before this one, and it is used to replace the sequence branch in the previous network.
The resulting model has exactly the same architecture and parameters, but with some weights set through training in a different task.
 
The performance of each auxiliary predictor is comparable to state-of-the-art methods, but due to possible overlaps between training and test sets of the different methods, a completely fair comparison would need more careful studies that are beyond the scope of this paper.

\subsection{Tricephalous network}
The ultimate application of model quality assessment is to rank models of the same protein.
In this section we will describe an architecture designed to exploit this structure of the problem by reducing it to pairwise comparisons.

\begin{figure}
	\centering
	\includegraphics[width=\linewidth]{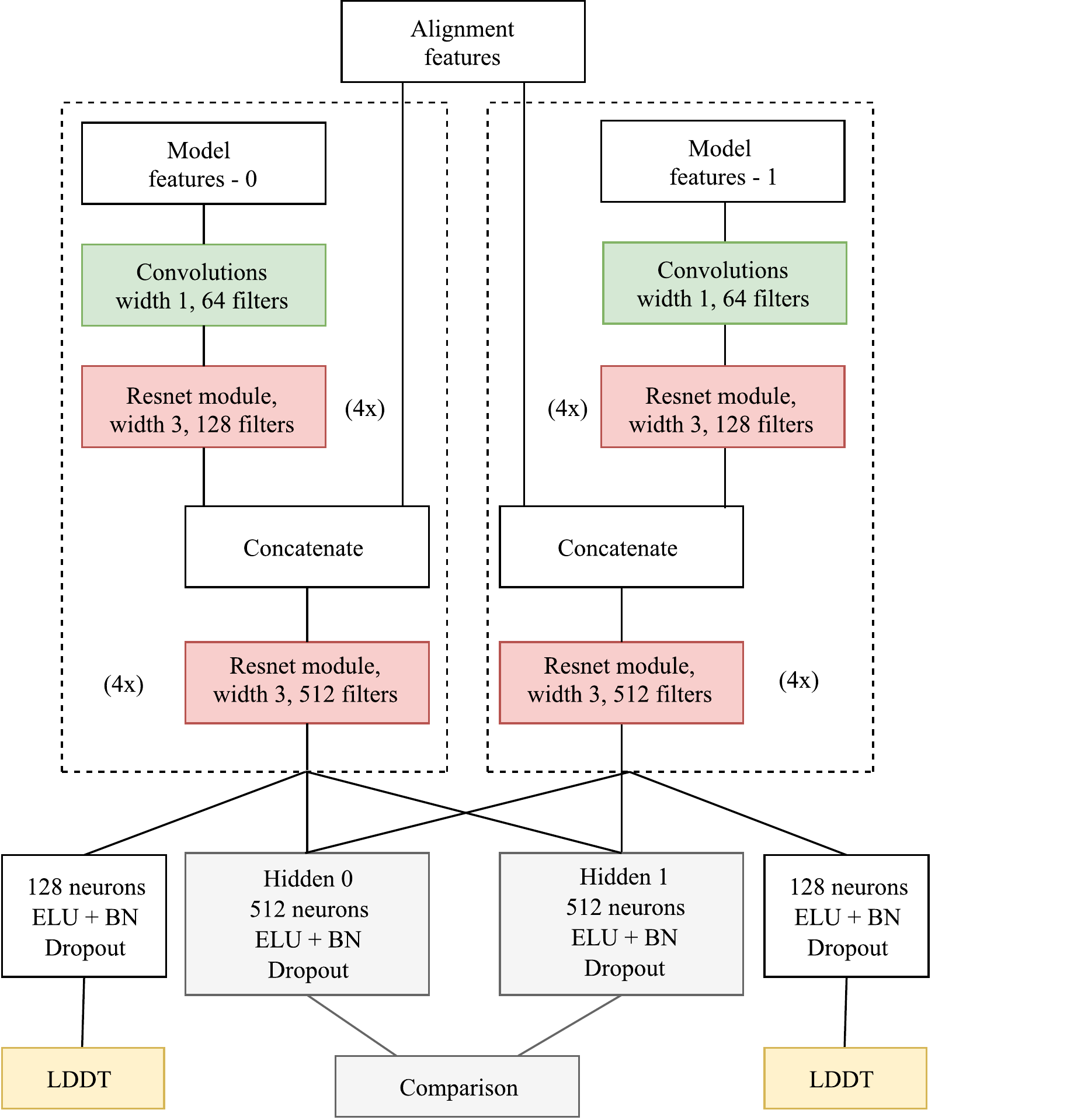}
	\caption{The Tricephalous architecture: the two stages of the Comparative network combined at once.}
	\label{fig:tric}
\end{figure}

In our dataset, for each target we have around \num{200} individual models, but \num{20000} pairs, which we can use as data augmentation.
So, instead of feeding one structure at the time, we will present the network with two models of the same target and ask it to predict the scores, as well as which model is better, for every residue, as shown in Figure~\ref{fig:tric}.

In order to ensure the symmetry of the problem is respected, each model is passed through a pair of identical copies of the pre-trained network described previously, keeping the parameters of each copy tied.
This is called a Siamese configuration, because the network is composed of two conjoined twins.
The comparison prediction is done by a symmetrised perceptron with one hidden layer, the SortNet~\citep{Sortnet}, represented in grey boxes in the Figure.
SortNet is a small variation of the classical perceptron to represent proper preference.
So, if we predict that $a$ is better than $b$ with a probability of $0.8$, we will also predict $b$ to be better than $a$ with a probability of $0.2$.
The SortNet is composed by two parallel hidden layers of 512 neurons per amino acid, and includes  batch normalisation and dropout~\citep{dropout}.

As is customary in multi-output models, the total loss is a weighted average of the loss for every output.
Since the comparison has effectively 200 times more training examples, we assign a higher weight on this output.
In our experiments, the best results were obtained with a weight of 1 for the comparison and $0.1$ for each of the scores.
See Figure~\ref{fig:tric}.

\subsection{Regression as a classification}
Deep learning usually performs better on classification than on regression tasks.
Therefore, we will divide the $[0, 1]$ range into $N$ equally sized bins and replace the mean squared error loss function by a cross entropy.
In order to recover the predicted score, we then average the scores with equally spaced weights.

\begin{equation}
p = \sum_{i=1}^{N} s_i  (\sigma_{low} + i \cdot \sigma_{step}) + \sigma_{offset},
\end{equation}
where $s_i$ is the predicted probability of being in the $i-th$ bin, $\sigma_{low}$, $\sigma_{step}$, and $\sigma_{offset}$ are three free parameters that were obtained minimising the mean squared error on the training set.

This is the final architecture, and we will refer to it as ProQ4.

\subsection{Dependencies}
The networks were implemented with Keras~\citep{keras}, using Tensorflow~\citep{tensorflow} as a backend. The data is stored in HDF5 files accessed through PyTables~\citep{pytables}.
All the networks were trained on a single Nvidia 1070Ti GPU, equipped with 8GB of VRAM.

In order to make predictions, the only dependencies are Python 3 with Numpy, Biopython, Keras, Tensorflow, and H5Py, DSSP, and a multiple sequence alignment.
All dependencies are open source and can be freely distributed.
Running on GPUs requires a CUDA-enabled NVIDIA GPU card, but can also work on CPUs.

\section{Results and discussion}
\begin{figure}
	\centering
	\includegraphics[width=\linewidth]{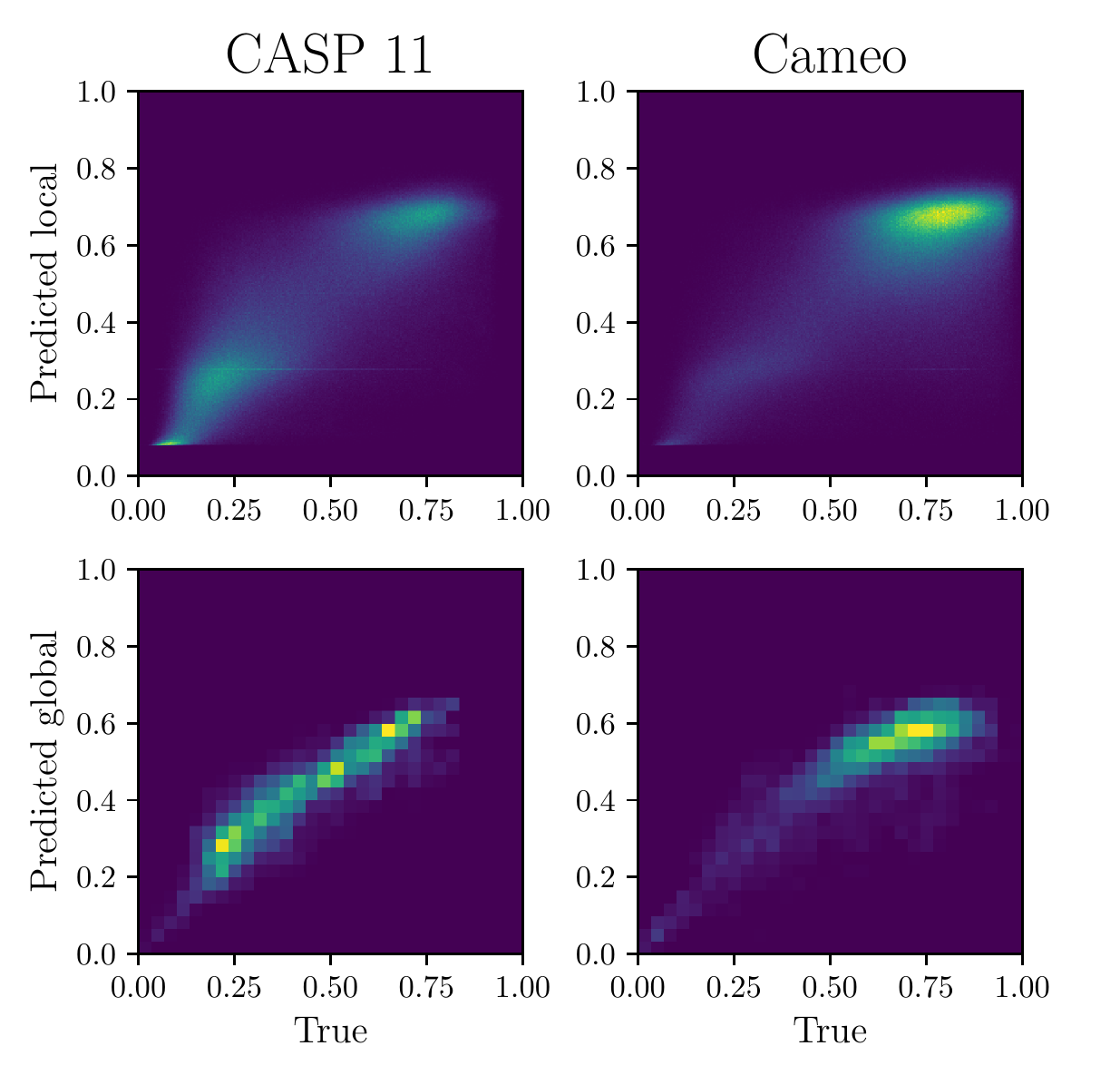}
	\caption{2D histogram of local (upper) and global (lower) scores for CASP 11 and Cameo.}
	\label{fig:loc_glob}
\end{figure}

In the Tables~\ref{tb:cc_local} and \ref{tb:cc_global} we present the results on the CASP 11 datasets for the three networks, compared with the state-of-the-art method, ProQ3D~\citep{ProQ3D}.
No other publicly available method was trained on LDDT, so we cannot establish a fair comparison.
We also present the results on Cameo~\citep{cameo} on Tables~\ref{tb:cc_local_cameo} and \ref{tb:cc_global_cameo}.
The last row on each table, Tricephalous network trained on classification is our final model, ProQ4, also plotted in Figure~\ref{fig:loc_glob}.

Of all the reported figures, we believe the per target correlation to be the most important metric, since one is usually interested in comparing models corresponding to the same protein.
First rank loss is also of interest, but since depends on a single data point per target, it is noisier.

\begin{figure}
	\centering
	\includegraphics[width=\linewidth]{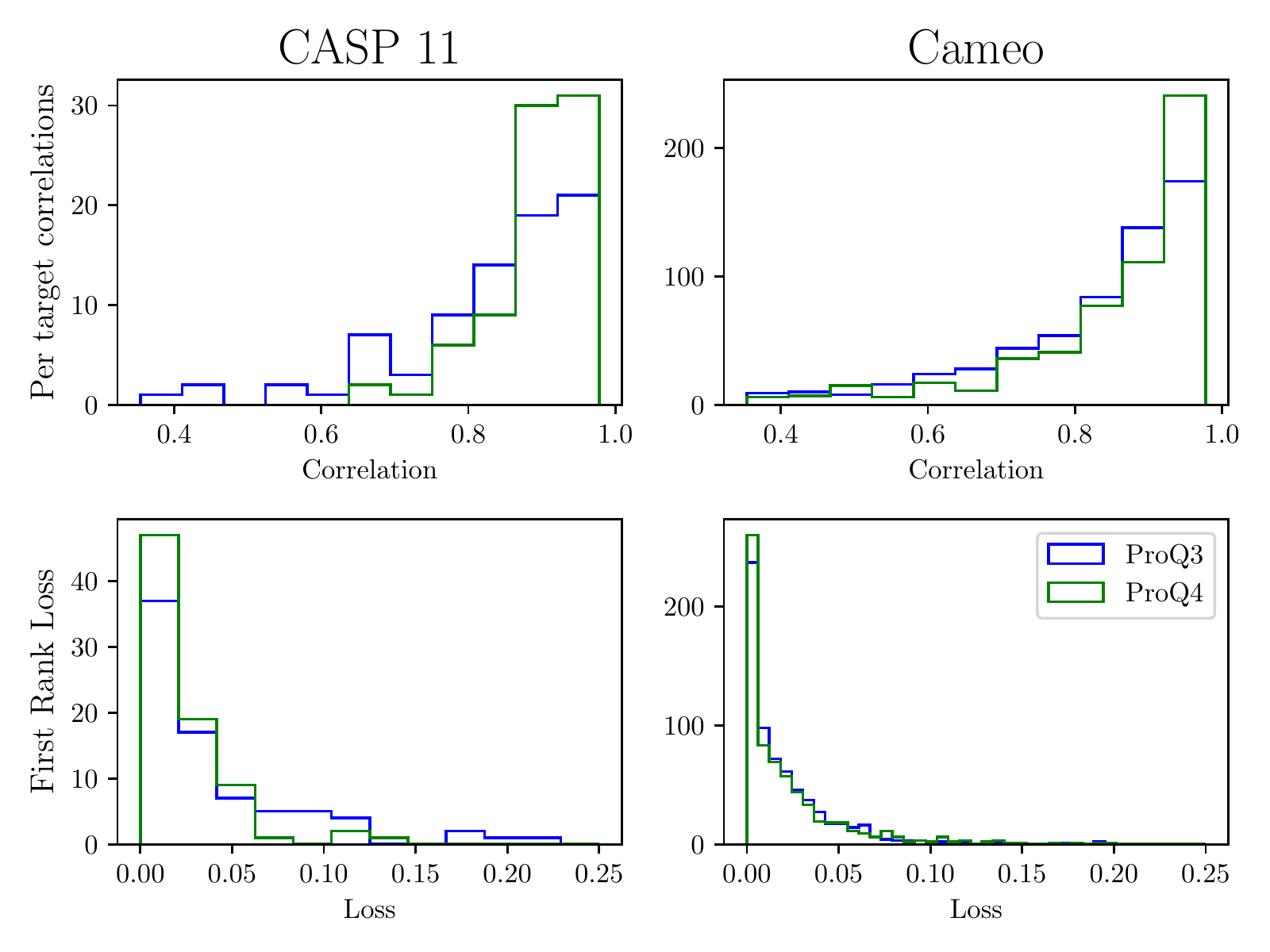}
	\caption{Comparison of the performance of ProQ3D (in blue) and ProQ4 (green) in per target correlations (upper) and first rank loss (lower).
		 For Cameo, the histograms have been truncated to the same range as CASP for clarity.}
	\label{fig:pertarget}
\end{figure}

For a baseline, we trained a simple feed forward network on our features with a window of 21 residues, with two hidden layers of 512 units, ELU activation, Batch Normalisation, and Dropout ($p_{drop}=0.1$).
This model represents how much information is there in the features themselves, disregarding the structure.

It can be seen that both the pre-training and the comparative strategy significantly improve the predictions, both the per target correlation and the first rank loss are better than previous methods.

\input{table_casp_2}
\input{table_cameo_2}

\begin{figure}[t]
	\centering
	\includegraphics[width=\linewidth]{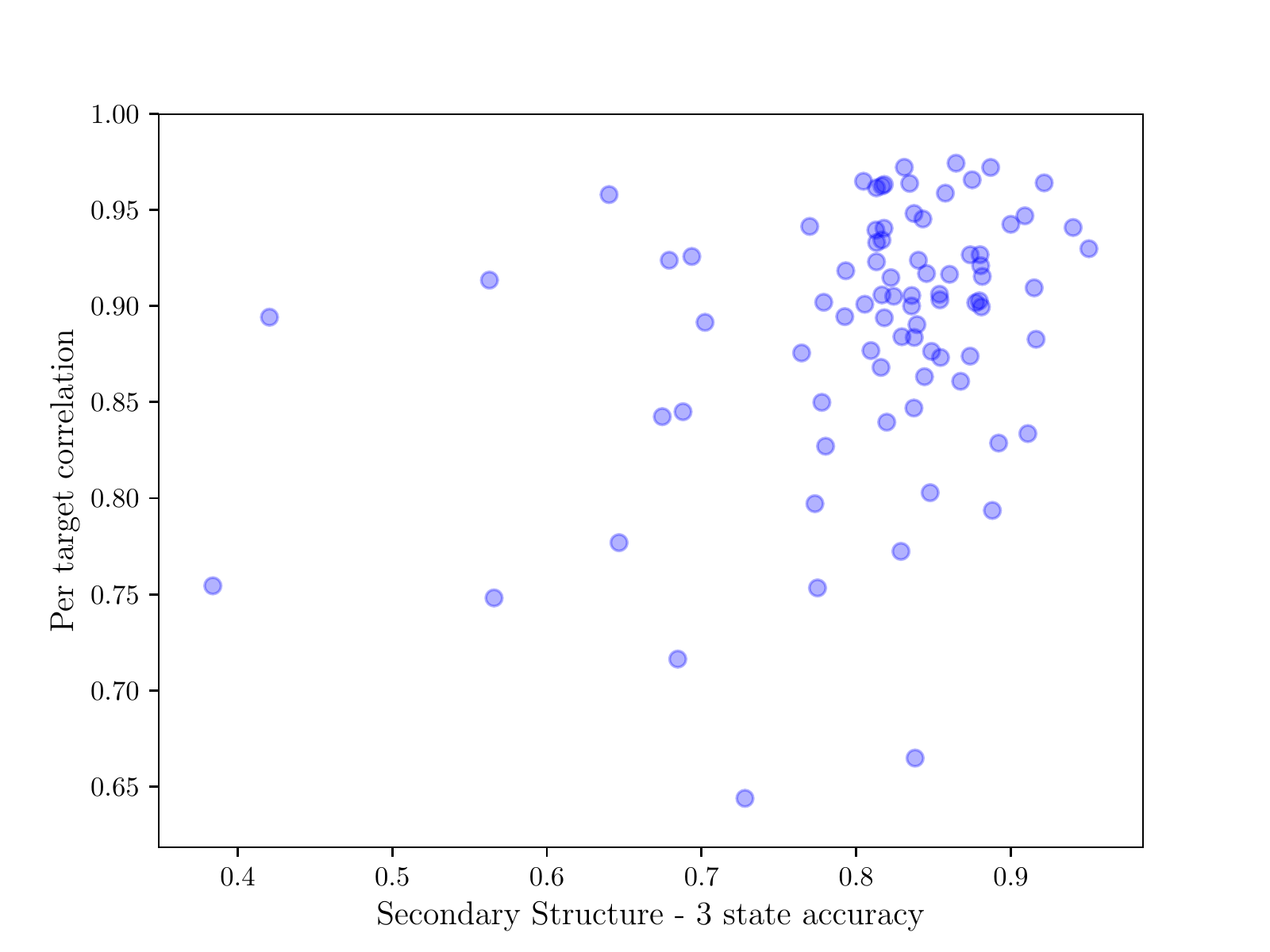}
	\caption{Comparison between per target correlations and the quality of the sequence-based predictor.
		The correlation is weak (0.33), suggesting that the quality of the sequence-based features is not a significant limiting factor.}
	\label{fig:ss_corr}
\end{figure}

\begin{figure}[t]
	\centering
	\includegraphics[width=\linewidth]{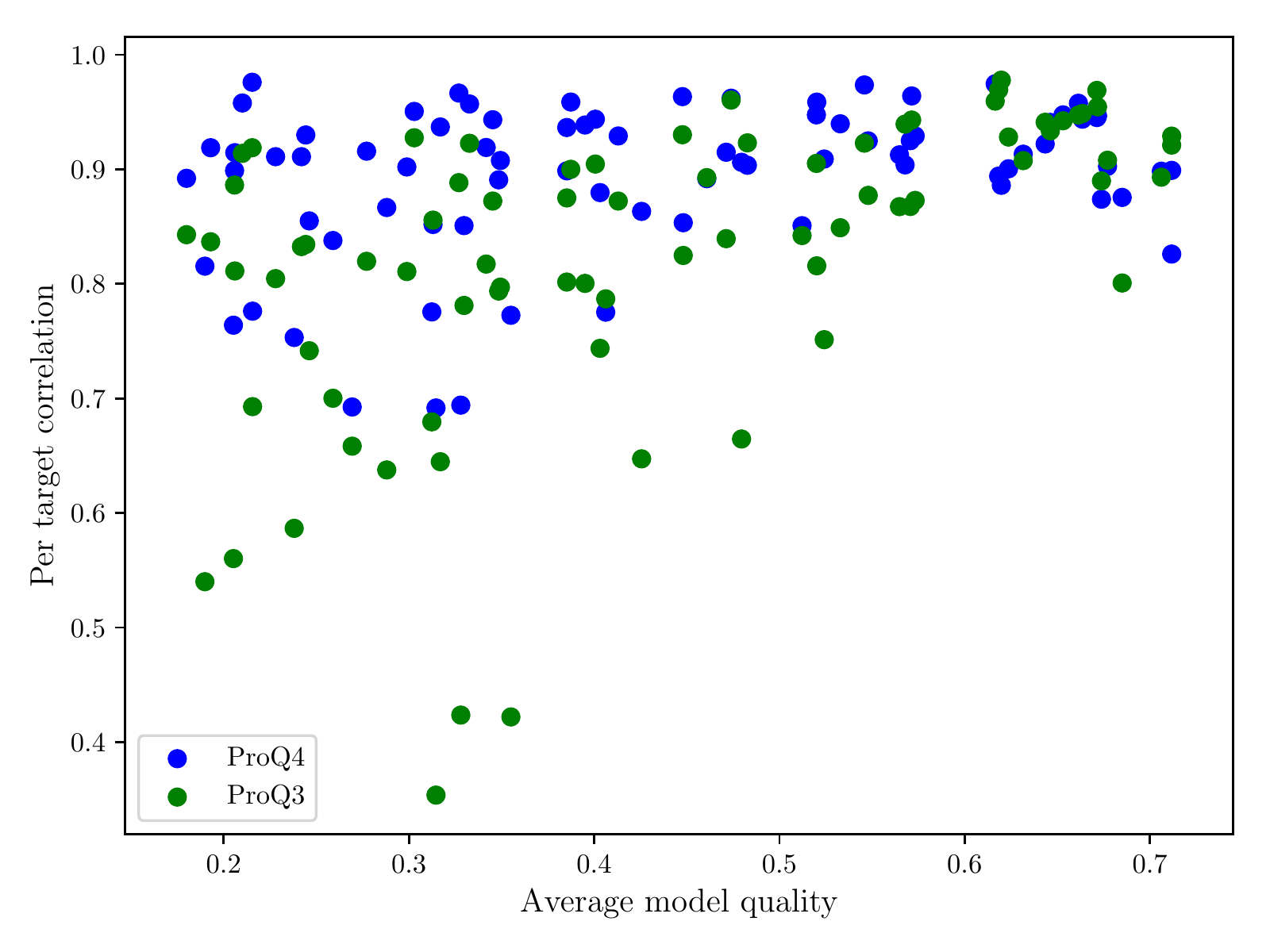}
	\caption{The per target correlations of ProQ3D and ProQ4 as a function of the average quality of the model on CASP11. The dependency is stronger for ProQ3D ($R_{PQ3D}=0.52$ vs $R_{PQ4}=0.31$).}
	\label{fig:quality_corr}
\end{figure}

The quality of the sequence-based network does not
appear to be hindering our performance, since the per target
correlations are only weakly correlated with the quality of the
predictions, see Figure~\ref{fig:ss_corr}.
Furthermore, the worst ranked targets have very low average scores, which are indeed difficult to rank, as shown in Figure~\ref{fig:quality_corr}.


\subsection{Number of classes in the scores}
When treating the regression as a classification, we tested all splits between 2 and 9 classes, using equally spaced bins.
The differences between 3 or more classes are smaller than $2\%$ in the local, global, and per target correlations.
The final model has 5 classes.

\subsection{The effect of larger networks}\label{sec:big_network}
We tested deeper networks increasing from 8 up to 50 ResNet blocks deep, and while it improved the local correlation, the global metrics remained the same, or slightly lower.
For example, with a depth of 50 layers, we increased a local correlation of from $0.77$ to $0.81$, on CASP11, while the global remained at $0.91$, but the per target drops from $0.90$ to $0.87$.
Even though the local correlations improve significantly, the per model correlation is slightly reduced from $0.56$ to $0.55$.
Larger networks seem to predict significantly better local scores, but being hampered by biases.

\subsection{Global vs local scores}
We are training on local scores, so directly trying to optimise the local RMSE.
It is surprising, then, that while the performance on local scores is significantly worse than ProQ3D, ProQ4 shows an improvement on global metrics, specially on per target correlations.
So, ProQ4 is better at telling the global differences between models, but fails to locate them accurately.
This may be due to the fact that during the training procedure we are presenting the network with the full model, or due to the lack of a description of the chemical properties of the model.

As shown in the section~\ref{sec:big_network}, an increase in local performance does not necessarily translate into global improvements.

\subsection{Difference between CASP and Cameo}
The main difference in Cameo is that there are much fewer models per target, and coming from fewer servers.
Many of the targets have easy templates available, so most of the models tend to be similar to each other.

The average first rank loss on the Cameo dataset is dominated by a single target, 4UYQ\_B, with a loss of 0.44 (see Figure~\ref{fig:pertarget}).
If that target were excluded, the average first rank loss would drop to 0.021, same as ProQ3D.

\subsection{Are we learning something new?}
\begin{figure*}[t]
	\centering
	\includegraphics[width=0.7\linewidth]{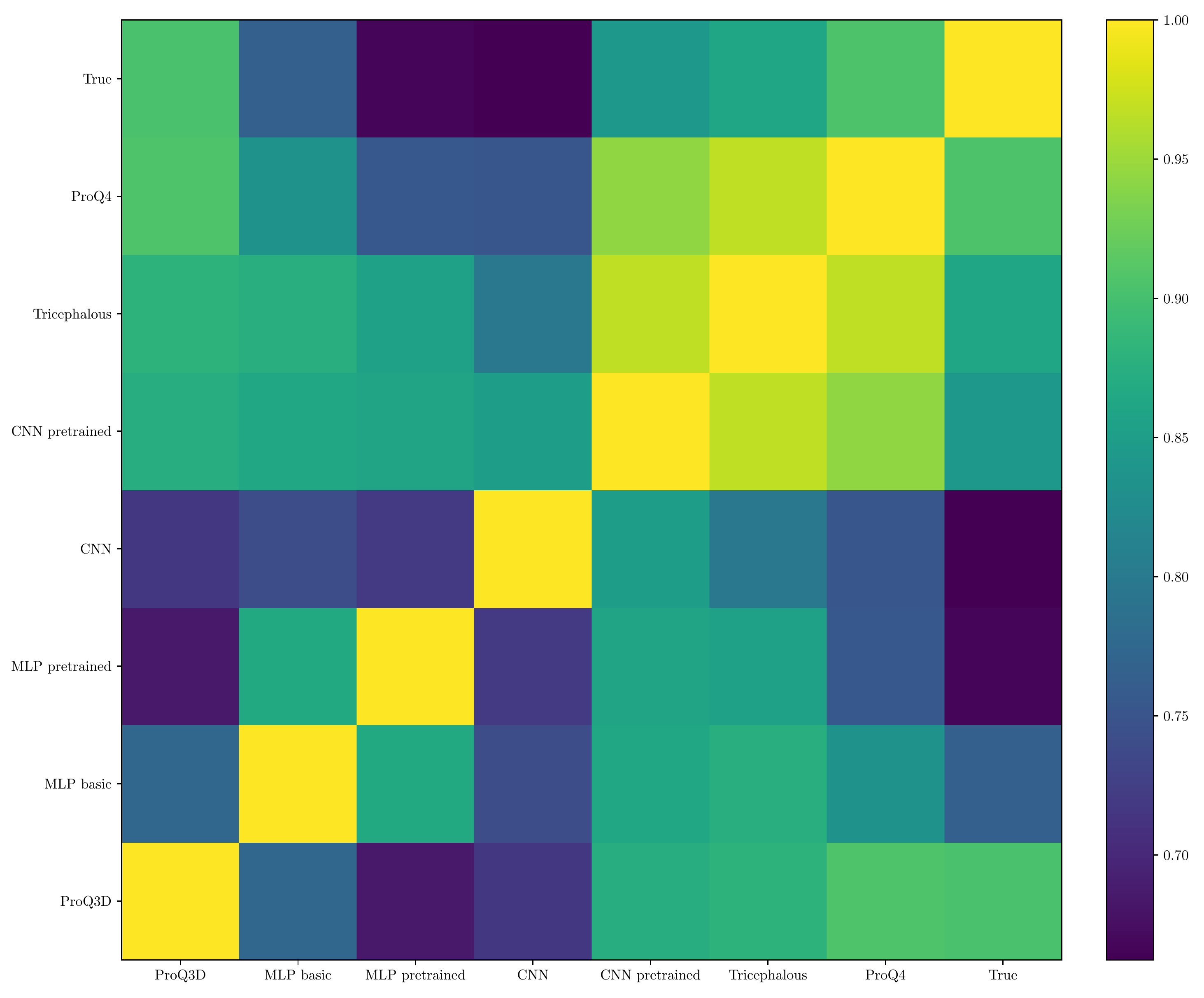}
	\caption{Correlation matrix between the predicted global scores on CASP 11 for each method.
		The order of the rows is the same as in the tables, with the last corresponding to the true scores.
	    A similar picture is obtained if we compare local scores instead.}
	\label{fig:corr_matrix}
\end{figure*}

Figure~\ref{fig:corr_matrix} shows a correlation matrix between all the global scores for all the architectures presented.
We can clearly identify two close clusters, one for the MLP architectures, and another one for all the convolutional networks using pre-trained features.
Simple CNN is an outlier, showing only a moderate correlation with its architectural twin, the pretrained network.
This indicates that the Tricephalous network and ProQ4 are learning the same thing, ProQ4 is just slightly better.

On the other hand, the agreement between ProQ3D and each of our models is similar to the agreement between the model and the true values, suggesting that both are learning something fundamentally different, despite having been trained on the same dataset.

\subsection{Method biases}
There are multiple strategies to generate protein models. Different methods fall into distinct kinds of
errors, and produce models of diverse chemical characteristics. This is a 
challenge when evaluating the model quality;  a model presenting unnatural
chemical properties can be due to low quality or just the existence of
non-optimised local geometries, for instance caused by sub-optimal
side-chain packing.

A common solution to the diversity problem is to uniformise the models by
repacking the side-chains with a single program such as
SCWRL~\citep{scwrl}. However, there are limitations to this
approach and it might be computationally expensive. Also, from a
practical point of view, it causes a method to be dependent on additional
programs.

In our proposed method, we use only a coarse description of the protein
that is not sensitive to the packing of the side chains. This should
make it quicker and easier to apply it in a large scale.

Another source of bias befalls when both the method generating the
model and the assessment make use of the same auxiliary predictor, such as PSIPRED~\citep{psipred}.
In this case, we find a bias for models generated with a particular
method. In order to tackle this issue, we replace the comparison with
explicit predictors with our own method and train on the hidden representation instead of the final output.

\section{Conclusions}
\subsection{Limitation of the method}
Our proposed method uses a very coarse description of the structural structures, focused on features that can be predicted from the sequence.
This restricted description limits the performance on the local level, but it is compensated by an increased performance in the global and, specially, per target ranking, due to the in built comparative structure.

Finding the right representation for the chemical properties of proteins from a deep learning point of view remains as an open question, and we hope this work will incentivise this line of research.


\subsection{The importance of structured data}
One of the reasons behind the great success of deep learning is the
ability to take full advantage of the structure of the data. For
machine learning in bioinformatics this has not been utilized
fully, except for contact prediction in the works of~\citet{marcin_nips} and~\citet{Wang28056090}.
Many machine learning methods in
bioinformatics still rely on sliding window approaches. Even when deep
learning has been applied this has often been limited to increasing
the complexity of a multi-layer perceptron architecture.

Since the advent of the Non Free Lunch theorem~\citep{no_free_lunch}, we know that the success of a machine learning algorithm is tied to how much domain knowledge can be included in its training.
In traditional machine learning, this is done through careful feature engineering, trying to find the closest representation to our objective.
In this work we propose a training framework that replaces the traditional end-to-end fitting with a multi-stage process designed to inject domain knowledge in every step of the way:
\begin{enumerate}
	\item Spatial relationships and translation invariance are coded as convolutions.
	\item The sequence information is extracted in the pre-training.
	\item The tricephalous architecture encodes the ranking nature of the problem.
\end{enumerate}

Both the pre-training and the comparative training bring an improvement to the results across all the metrics we have evaluated on both datasets.

The importance of structure is also seen on the effect of pre-training.
While it improves CNN-based architectures, Multi Layer Perceptron (MLP) models don't benefit, or are even hindered by pre-training.



\section*{Funding}
This work was supported by grants from the Swedish Research Council
(VR-NT 2016-03798 to AE) and Swedish e-Science
Research Center (BW). The Swedish National Infrastructure provided computational resources for Computing (SNIC) at NSC.\vspace*{-12pt}

\FloatBarrier

\bibliography{refs2}

\end{document}

%% file: table_casp_2.tex
\begin{table*}[h!]
	\centering
		\caption{Summary of results on the local scores on CASP11.\label{tb:cc_local}}
				\begin{tabular}{lccc}
					\toprule
					Method & R-local  & RMSE local & R-per model\\
					\midrule
					ProQ3D \\ (retrained on LDDT) &  \textbf{0.84} & \textbf{0.125} & \textbf{0.61} \\
					\midrule
					Base MLP  & 0.62 & 0.180 & 0.44 \\
					Pre-trained MLP  & 0.52 &  0.198 & 0.47 \\
					\midrule
					Simple CNN & 0.51 & 0.205 &0.42\\
				    Pre-trained CNN  & 0.68	 & 0.172 & 0.55 \\
					Tricephalous & 0.72 & 0.160 & 0.55\\
					\midrule
				    ProQ4 & 0.77 & 0.147 & 0.56 \\
					\bottomrule
				\end{tabular}
			\caption*{R-local is the correlation between all local predicted and true scores in the dataset; RMSE stands for Root Mean Squared Error. R-per model is the average correlation between predicted and true scores for each model in the dataset. Find a more detailed explanation in section~\ref{sec:metrics}}\medskip
		\caption{Summary of results on the global scores on CASP11.\label{tb:cc_global}}{
			\begin{tabular}{lcccc}
				\toprule
				Method   & R-global & Global RMSE & R-per target & First rank loss \\
				\midrule
				ProQ3D\\(retrained on LDDT) & 0.90 & 0.080 & 0.82 &  0.040\\
				\midrule
				Base MLP  & 0.77 & 0.118 & 0.82 & 0.046\\
				Pre-trained MLP  & 0.67 & 0.135 & 0.82 & 0.046\\
				\midrule
				Simple CNN & 0.67 & 0.157  & 0.68 & 0.091\\
				Pre-trained CNN & 0.85 & 0.121  & 0.85 & 0.036 \\
				Tricephalous& 0.87 &  0.106 & 0.88 & 0.029 \\
				\midrule
				ProQ4  & \textbf{0.91} & \textbf{0.085}  & \textbf{0.90} & \textbf{0.022}\\
				\bottomrule
			\end{tabular}}
		\caption*{R-global is the correlation between all global predicted and true scores in the dataset; R-per target is the average correlation of global scores of models for each protein; and first rank loss is the average difference in true scores between the best model and the top ranked for each target.}

\end{table*}

%% file: table_cameo_2.tex
\begin{table*}[h!]
    \centering
		\caption{Summary of results on the local scores on Cameo.\label{tb:cc_local_cameo}}{
				\begin{tabular}{lccc}
					\toprule
					Method & R-local  & RMSE local & R-per model\\
					\midrule
					ProQ3D\\(retrained on LDDT) &  \textbf{0.79} & \textbf{0.137} & \textbf{0.64} \\
					\midrule
					Base MLP  & 0.48 & 0.232 & 0.42 \\
					Pre-trained MLP  & 0.50 &  0.248 & 0.47 \\
					\midrule
					Simple CNN & 0.43 & 0.219 & 0.41\\
					Pre-trained CNN  & 0.61	 & 0.202 & 0.55 \\
					Tricephalous & 0.65 & 0.199 & 0.56\\
					\midrule
					ProQ4 & 0.65 & 0.201 & 0.56 \\
					\bottomrule
				\end{tabular}
			}
		\caption*{R-local is the correlation between all local predicted and true scores in the dataset; RMSE stands for Root Mean Squared Error. R-per model is the average correlation between predicted and true scores for each model in the dataset. Find a more detailed explanation in section~\ref{sec:metrics}}\medskip

		\caption{Summary of results on the global scores on Cameo.\label{tb:cc_global_cameo}}{
			\begin{tabular}{lcccc}
				\toprule
				Method   & R-global & Global RMSE & R-per target & First rank loss \\
				\midrule
				ProQ3D\\(retrained on LDDT) & \textbf{0.88} & \textbf{0.093} & 0.80 &  \textbf{0.021}\\
				\midrule
				Base MLP  & 0.71 & 0.180 & 0.75 & 0.034\\
				Pre-trained MLP  & 0.76 & 0.202 & 0.76 & 0.032\\
				\midrule
				Simple CNN & 0.70 & 0.163 & 0.71 & 0.041\\
				Pre-trained CNN & 0.80 & 0.158  & 0.80 & 0.028 \\
				Tricephalous& 0.82 &  0.158 & 0.83 & 0.025 \\
				\midrule
				ProQ4  & 0.82 & 0.156  & \textbf{0.84} & 0.023\\
				\bottomrule
			\end{tabular}
		}
	\caption*{R-global is the correlation between all global predicted and true scores in the dataset; R-per target is the average correlation of global scores of models for each protein; and first rank loss is the average difference in true scores between the best model and the top ranked for each target.}

\end{table*}